# Extending Integrated Assessment Model scenarios until 2150 using an emulation Framework

Weiwei Xiong[1,*], Katsumasa Tanaka[1,2,*]

1 Laboratoire des Sciences du Climat et de l'Environnement (LSCE), IPSL, CEA/CNRS/UVSQ, Université Paris-Saclay, Gif-sur-Yvette, 91191, France

2 Earth System Division, National Institute for Environmental Studies (NIES), Tsukuba, 305-8506, Japan

*These authors contributed equally.

Corresponding author: Weiwei Xiong (weiwei.xiong@lsce.ipsl.fr), Katsumasa Tanaka (katsumasa.tanaka@lsce.ipsl.fr)

**Abstract:** Whereas there is growing interest in exploring longer-term climate with overshoot, including tipping elements, beyond 2100, most Integrated Assessment Models (IAMs) generate emissions scenarios only till 2100. Here we propose a framework to extend scenarios until 2150 using an emulator of IAMs. Our framework offers a potential interim solution for developing very long-term scenarios, such as the Scenario Model Intercomparison Project (ScenarioMIP), circumventing the challenges of fully simulating IAMs beyond 2100.

**Keywords:** Long-term scenario; Scenario extension; IAM emulator

## 1 Introduction

Process-based IAMs are widely used for projecting future climate scenarios and evaluating mitigation policies (Weyant 2017). These models produce diverse greenhouse gas (GHG) emission trajectories based on various socio-economic and technological assumptions (Riahi *et al* 2017). Such scenarios have enabled assessments of climate mitigation strategies through the end of the 21st century (O'Neill *et al* 2020) – a time horizon that has remained largely unchanged since the 1990s (Houghton *et al* 1990, Nakićenović and Swart 2000). On the other hand, several climate science research communities call for scenarios beyond the 21st century to explore longer-term climate dynamics with overshoot, including tipping points (Ritchie *et al* 2021, Armstrong McKay *et al* 2022). Projections into the 22nd century are essential for assessing future climate risks and designing robust policy interventions (Lyon *et al* 2022, Easterling *et al* 2024).

To bridge this gap, idealized long-term scenarios beyond 2100 have been proposed and widely applied (Meinshausen *et al* 2020, van Vuuren *et al* 2025). However, these extended scenarios typically rely on simplified linear assumptions, such as holding emissions constant after 2100 or linearly converging to a specified target such as net-zero. Such assumptions do not necessarily reflect the socio-economic and technological dynamics represented in IAMs until 2100. These limitations undermine the coherence between pre-2100 scenarios and their extensions.

## 2 Methods

### 2.1 The emulator of IAMs

Here we present a methodological framework that extends emissions scenarios until 2150 using an emulator of IAMs (emIAM) (Xiong *et al* 2025, hereafter, X25; see Appendix A1). Our emulator captures the socio-economic and technological dynamics of the underlying IAM and maintains them beyond 2100 in the extended scenarios, ensuring that the extended scenarios remain consistent with the original pre-2100 scenarios. In situations where it is challenging to simulate IAMs beyond 2100, as observed in the Scenario Model Intercomparison Project (ScenarioMIP) for the Coupled Model Intercomparison Project, Phase 7 (CMIP7) (van Vuuren *et al* 2025), our framework offers a potential interim solution until all technical requirements are met for simulating such IAMs over longer time horizons. This article focuses on process-based IAMs typically used in a cost-effectiveness setting for developing detailed mitigation scenarios. While cost-benefit IAMs run over multiple centuries to generate mitigation scenarios, they usually do not produce pathways aimed at specific climate targets. We also acknowledge that there are a few exceptional process-based IAMs that simulate beyond 2100 (Azar *et al* 2013). The framework we propose here is for standardized scenario extension across multiple IAMs up to 2150, whereas a more detailed, IAM-specific approach is explored elsewhere (Tanaka *et al* 2025).

To demonstrate how our emulator framework works for extending IAM scenarios through 2150, we initially consider a set of nine IAMs included in the open-access ENGAGE Scenario Explorer (part of the IPCC AR6 Scenario Explorer and Database). We chose the ENGAGE database because scenarios simulated by these IAMs are available for a range of cumulative carbon budgets. Using those scenarios, X25 developed emIAM representing each IAM. The emIAM's core element is the time-invariant marginal abatement cost (MAC) curve, which aims to capture the relationship between carbon price (or $CO_2$-equivalent price using the 100-year Global Warming Potential (GWP100)) (Tanaka *et al* 2010) and emission abatement (in percentage relative to the baseline scenario) of each GHG ($CO_2$, $CH_4$, and $N_2O$) in each IAM until 2100, which we term the price-quantity relationship here. Although $SO_2$ is also an important gas for determining future climate, we did not apply the emulator to $SO_2$ because $SO_2$ emissions are substantially influenced by clean air policies assumed in IAMs, largely independent of the carbon price. Such price-quantity relationships are not necessarily time-invariant due to capital stock, constraints on technology expansion, and the availability of new technologies. Nevertheless, X25 shows that time-invariant MAC curves adequately capture the price-quantity relationships at least for a subset of IAMs in the latter half of this century. This provides a basis for applying our emulator to selected six IAMs (AIM/CGE v2.2, GEM-E3 v2021, MESSAGEix-GLOBIOM 1.1, POLES-JRC ENGAGE, REMIND-MAgPIE 2.1-4.2, and WITCH 5.0) for extending the scenarios for another 50 years. The use of time-invariant MAC curves allows us to develop mitigation pathways beyond 2100 without needing to infer new MAC curves after 2100.

**2.2 Scenario extension approaches**

We extend three arbitrarily selected scenarios from each IAM with carbon budgets of 600, 1000, and 1400 $GtCO_2$ during 2019 – 2100. As the first step, we extend the baseline scenario (i.e., EN_NPi2100 in ENGAGE) until 2150 for each gas from each IAM by linearly extrapolating the respective baseline emission trend from the previous 20 years (2080-2100). We then extend the three mitigation scenarios of each IAM using the following four distinct Approaches.

1. *Extending carbon price pathways by linear extrapolation to 2150 using the 2090–2100 trend (Price_Linear)*: The extended carbon price pathways uniquely determine the post-2100 emission levels though respective MAC curves. See Appendix A2 for the rationale for using the 2090-2100 period as a basis for extrapolation. To ensure data continuity, the pre-2100 carbon price pathways are slightly adjusted using scaling such that the resulting 2100 emissions implied by MAC curves match the corresponding 2100 emissions from the respective IAMs.

2. *Extending carbon price pathways by linear convergence to the "net-zero" carbon price by 2150 (Price_ZeroEmi)*: The carbon price corresponding to net-zero $CO_2$ emissions can be obtained from the $CO_2$ MAC curve of each IAM. The abovementioned adjustment of carbon price pathways is also applied.
3. *Minimizing the net present value of the total cost of abating GHG emissions for a given carbon budget target by 2150 (CumBudg_Opt)*: The emission and carbon price pathways are optimized under the 2100-2150 carbon budgets (-400, -500, and -600/-550 $GtCO_2$) (see Appendix A2).
4. *Minimizing the net present value of the total cost of abating GHG emissions for a given temperature target by 2150 (T2M_Opt)*: The emission and carbon price pathways are optimized for achieving the temperature target of 1.5°C by 2150 with overshoot. For the temperature calculations, emIAM is directly coupled with the climate emulator, ACC2 (Tanaka *et al* 2007, Tanaka and O'Neill 2018; see Appendix A1).

In all four Approaches, the post-2100 emissions of GHGs and air pollutants that are not included in emIAM are held constant at their 2100 levels. It is important to note that Approach 3 is not technically consistent with the IAMs that do not perform inter-temporal optimizations (i.e., AIM/CGE V2.2, GEM-E3 V2021, and POLES-JRC ENGAGE). Approach 4 goes beyond how the IAMs operate, as the IAM and climate emulators fully interact in the optimization, without relying on the remaining carbon budget approach directly or indirectly used in these IAMs.

## 3 Results

Taking REMIND-MAgPIE 2.1-4.2 (Kriegler *et al* 2017) as an example, we present the extended $CO_2$, $CH_4$, and $N_2O$ emission trajectories based on the four Approaches (Fig. 1). For $CO_2$ emissions, Approaches 1 and 2 using the carbon price as the basis of extrapolation provide smooth emissions extensions to 2150. In contrast, Approaches 3 and 4 yield scenarios with more complex profiles. In an intertemporal optimization framework, an optimal solution calculated over 2019–2100 would generally differ from one optimized over the full 2019–2150 horizon in a single run. This discrepancy is exemplified by initial emission rebounds followed by deep mitigation, as a result of inter-temporal cost minimizations for the specific carbon budget or temperature target. These are subsequently followed by a rise in emissions shortly before 2150, which is an artifact of the target-setting mechanism, where aggressive emission reductions become mathematically unnecessary just before the target year. For $CH_4$ and $N_2O$ emissions, all extended pathways are similar, showing a steady decline in most cases.

Approaches 1 and 2 based on *price* extrapolation offer the following two advantages over the conventional approach based on *emission* extrapolation.

- Provide a mathematically stable basis for extrapolation using a monotonic trend of carbon price pathways
- Incorporate the pre-2100 socio-economic and technological dynamics captured by MAC curves (under certain assumptions as discussed earlier)

On the other hand, the conventional emission-based approach offers the following two advantages.

- Computationally lighter, requiring no IAM emulator
- Allow for a more straightforward incorporation of an emission target in the post-2100 period

Regarding Approaches 3 and 4, splitting the inter-temporal optimization across 2100 indeed creates an artificial boundary effect. Such emission profiles are usually not desirable in extended scenarios. While Approaches 3 and 4 offer different applications elsewhere (X25, Tanaka *et al* 2025), they require additional assumptions (e.g., carbon budget or temperature target for the post-2100 period). Hence, they are not suited for standardized scenario extension for multiple IAMs up to 2150.

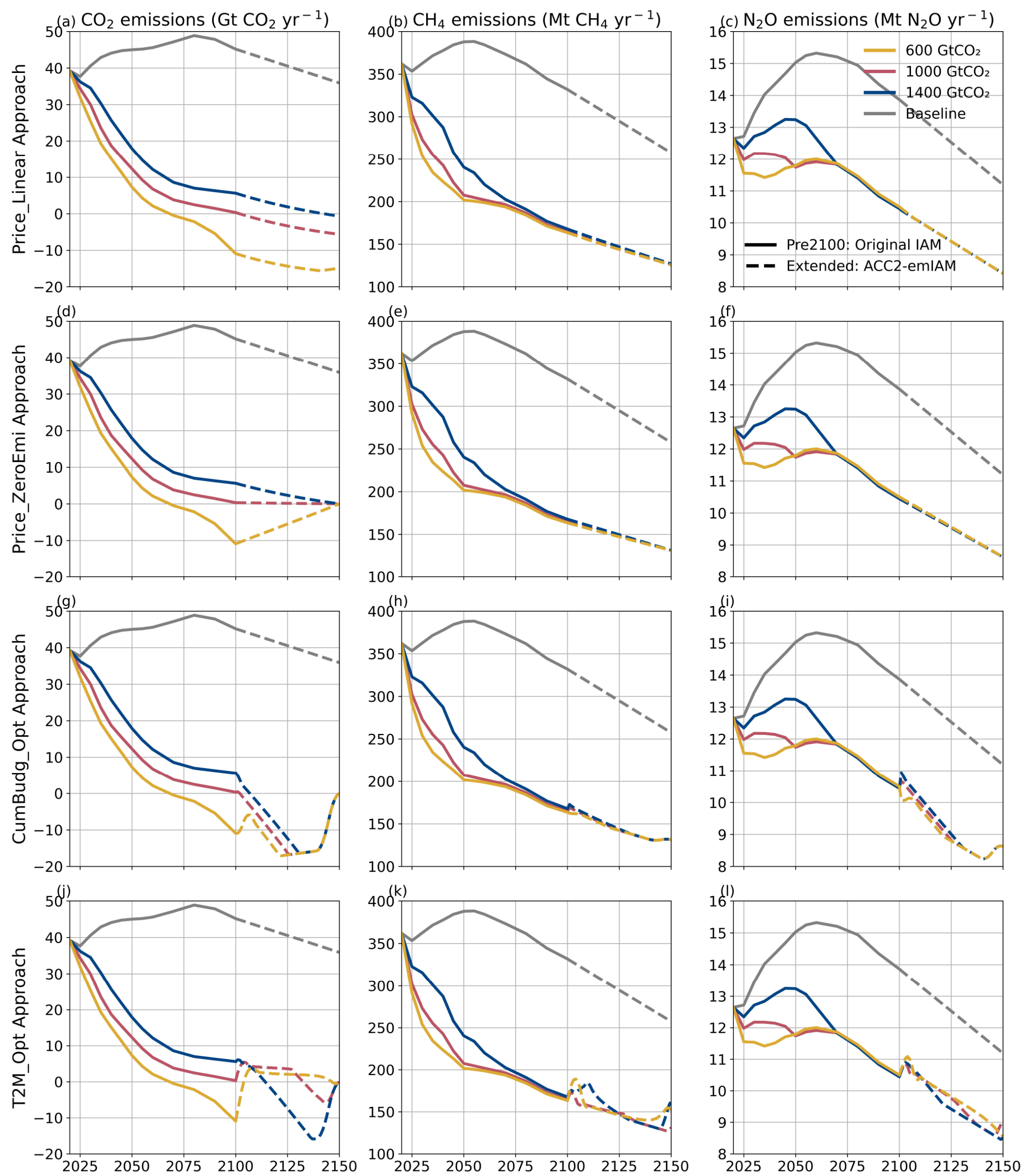

**Figure 1. Multi-gas emission pathways developed by REMIND-MAgPIE 2.1-4.2 with extension to 2150 using four different Approaches.** Each column corresponds to a gas ($CO_2$, $CH_4$, and $N_2O$), and each row corresponds to an Approach (*Price_Linear*, *Price_ZeroEmi*, *CumBudg_Opt*, and *T2M_Opt*). Gray lines indicate the baseline scenario; colored curves represent mitigation scenarios with different carbon budgets. Solid lines depict the original IAM results through 2100 and dashed lines the 2100–2150 extension.

Among the four Approaches, the price-based Approaches 1 and 2 are, therefore, the most promising alternatives to the conventional emission-based approach. We now present the results from these two Approaches for all six IAMs (Fig. 2). Both provide smoothly and continuously extended scenarios in all cases. With Approach 1, all emission scenarios are extended to a further lower level, except for those from AIM/CGE v2.2, which exhibits an emission rebound due to the rapidly declining trend of carbon prices between 2090 and 2100.

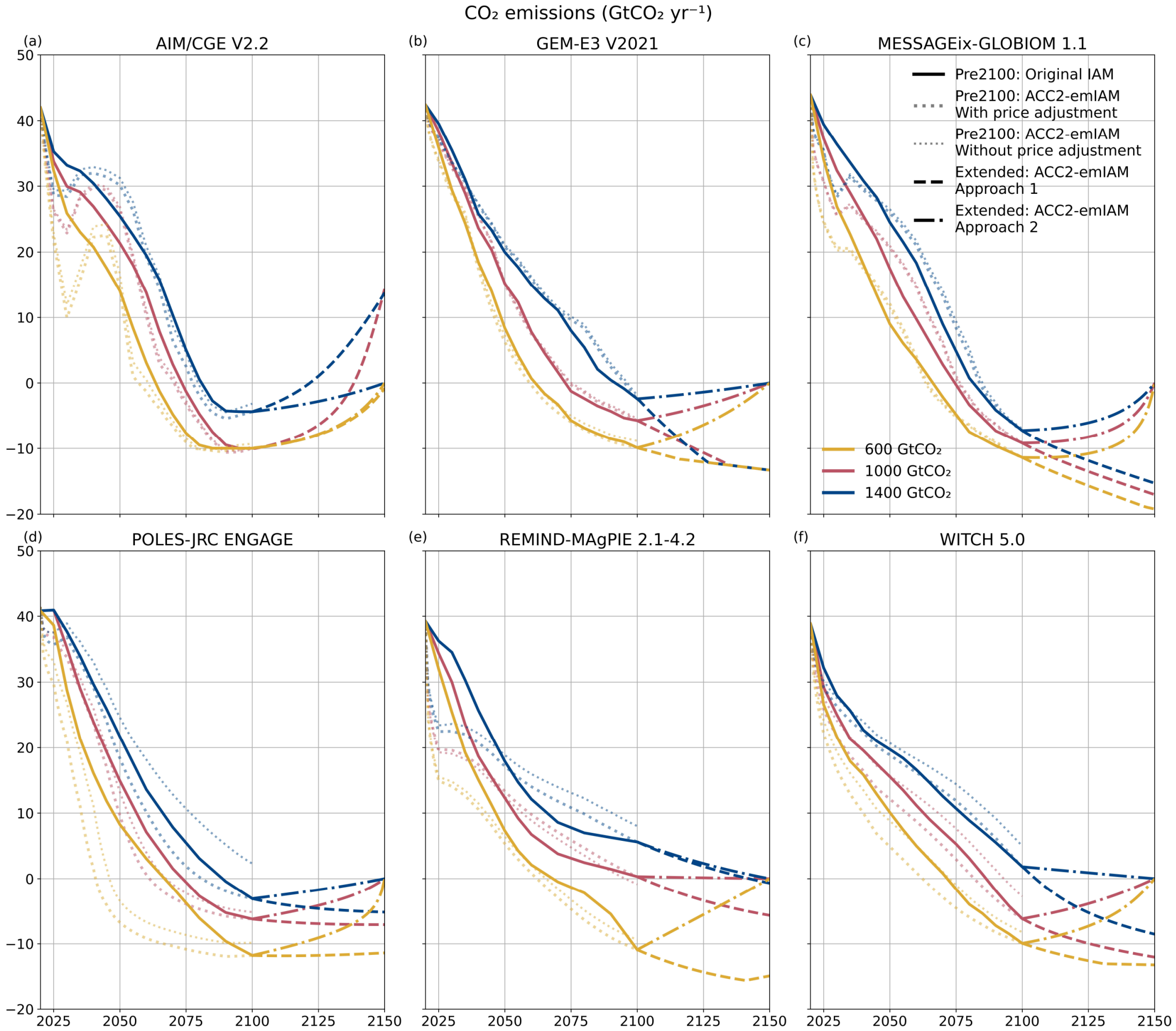

**Figure 2. $CO_2$ emission pathways developed by six IAMs with extension to 2150 using Approaches 1 and 2.** Colored lines represent carbon budget scenarios. Solid lines depict the original IAM scenarios through 2100; thick and thin dotted lines show the pre-2100 scenarios calculated by our emulator with and without the adjustment for carbon price pathways, respectively; dashed and dashed dotted lines are the extended scenarios until 2150 based on Approaches 1 and 2, respectively.

To validate Approaches 1 and 2, we further reproduce the emission pathways until 2100 that are inversely calculated from the associated MAC curves by prescribing the IAM's carbon price pathways (Fig. 2). Deviations from the original IAM emission pathways result from the misfit of the MAC curves to the original IAM data. While our time-invariant MAC curves only approximately capture the complex price–quantity relationship in IAMs, the deviations are smaller in the second half of this century than in the first half as shown by X25, supporting the use of Approaches 1 and 2 for extending scenarios. Additional validation results are provided in the Supplementary (Figs. S3 and S4 for Approaches 1 and 2; Figs. S6–S8 and S10-S12 for Approaches 3 and 4). For a comprehensive discussion of these validation outcomes, see X25.

**4 Discussion and conclusions**

To apply this framework to latest IAMs, the IAM emulator needs to be re-calibrated, a task requiring at least six to seven scenarios from each new IAM at a range of mitigation levels. While this paper focuses on global total emissions, our framework can also be applied to sectoral and regional emissions. The spatial patterns of emissions are beyond the scope of our current emulation but can be assumed to be fixed after 2100 as a first approximation.

Overall, the proposed emulator-based framework can extend diverse emission pathways to 2150, while retaining the fundamental price-quantity relationships of original IAMs. The choice of extension approach depends on its research objective, such as prioritizing computational simplicity versus preserving underlying IAM dynamics. Introducing our price-based alternatives expands the extension approaches available to the community. Our framework can complement existing extended scenarios based on linear emission extrapolations and may serve as an interim solution until process-based IAMs can become fully operational for longer simulations beyond 2100.

## Appendix

### A1 Additional details on the IAM emulator

We employ the recently developed IAM emulator, emIAM v1.0 (X25), which builds on a MAC curve approach that aims to capture the relationship between carbon price and emission abatement of each IAM. The MAC curves are represented by the function $f(x) = ax^b + cx^d$, where $f(x)$ denotes the carbon price relative to baseline, $x$ is the abatement level in percentage relative to baseline, and $a, b, c, d$ are the parameters specifically calibrated to each IAM. Furthermore, using constraints on the first- and second-derivatives of gas-specific abatement rates (also calibrated to each IAM), our emulator approximately represents technological change and socio-economic inertia.

emIAM v1.0 was developed to emulate multiple IAMs in the ENGAGE Scenario Explorer (Riahi *et al* 2021, Drouet *et al* 2021). This database provides a comprehensive set of GHG emission pathways generated by various IAMs under different carbon budget constraints. The IAM emulator captures the price-quantity relationship in those IAMs using an extensive set of MAC curves at multiple levels (global, regional, and sectoral), although this paper focuses on the global MAC curves for three gases, $CO_2$, $CH_4$, and $N_2O$.

emIAM covers nine widely used IAMs—AIM/CGE v2.2, COFFEE 1.1, GEM-E3 v2021, IMAGE 3.0, MESSAGEix-GLOBIOM 1.1, POLES-JRC ENGAGE, REMIND-MAgPIE 2.1-4.2, TIAM-ECN 1.1, and WITCH 5.0—many of which serve as providers of marker scenarios for ScenarioMIP. emIAM v1.0 has been validated to reproduce the original IAMs' GHG emissions in many different ways (Chapter 4 of X25). In our scenario extension exercise, IMAGE 3.0 and TIAM-ECN 1.1 have been excluded because the projections of these IAMs were not well reproduced by our emulator (Table S1). In addition, COFFEE 1.1 has been also excluded because the underlying $N_2O$ MAC curve did not adequately capture the underlying price-quantity relationship (Fig. S2 of X25).

In addition, to extend emission pathways with temperature goals (i.e., *T2M_Opt* Approach), we coupled the IAM emulator with the reduced-complexity climate model ACC2 (Tanaka *et al* 2007, Tanaka and O'Neill 2018), hereafter ACC2-emIAM. ACC2 represents primarily important Earth system processes at the global-mean level through a carbon cycle box model, highly parameterized atmospheric chemistry module, and heat-diffusion climate module, covering a wide range of climate forcers (GHGs and aerosols) with key nonlinear feedback such as $CO_2$ fertilization and climate carbon-cycle feedback. ACC2-emIAM is able to produce cost-effective emission pathways that comply with specified climate targets within a single modeling framework.

## A2 Additional details on the scenario extension approaches

We consider three distinctive scenarios (i.e., scenarios with the cumulative carbon budget of 600, 1000, and 1400 $GtCO_2$) from each IAM for scenario extension. To extend these IAM emission pathways beyond 2100, we developed the following four Approaches (see Fig. S1):

1. *Price_Linear* Approach. The carbon price is linearly extrapolated to 2150 from its 2090–2100 trend. Most models exhibit a monotonic trend in the carbon price pathway, except for AIM/CGE v2.2, which shows a price peak in 2090, followed by a decline (Fig. S2). To ensure the carbon price extension in a representative way over a long term, we use the 2090-2100 trend as the basis for extrapolation.

2. *Price_ZeroEmi* Approach. We set the 2150 carbon price at the level corresponding to 100% $CO_2$ abatement. We apply a linear interpolation between the carbon price in 2100 and the "zero-emission" price in 2150.

Both *Price_Linear* and *Price_ZeroEmi* are carbon price-based approaches. Carbon pricing is one of the core drivers that determine the abatement allocation in IAMs. Cumulative emission constraints are sometimes converted to corresponding carbon price pathways, depending on the type of input required for IAMs. In these approaches, carbon price acts as the primary input in our emulator; once the price pathway is set, the abatement levels of each gas are uniquely determined accordingly. Constraints on the first- and second-derivatives of abatement levels are thus not used. However, because emIAM does not fully capture the IAM's price–quantity relationships, emissions projected by these carbon price-based approaches do not exactly match the original IAM emissions in 2100. Therefore, we introduced a gas-specific scaling factor to adjust the carbon price pathway of each IAM so that it matches the original emissions in 2100.

3. *CumBudg_Opt* Approach. This is the cumulative carbon budget-based approach. Building on the IAM practice of generating scenarios under various cumulative carbon budget constraints, we apply this approach to the post-2100 period, while maintaining the same emission pathways as the original IAMs up to 2100. In the absence of a benchmark for allocating budgets up to 2150, we hypothetically assign differentiated 2100-2150 targets of –600/–550 (where –550 $GtCO_2$ adopted when the –600 $GtCO_2$ run is infeasible because of the limitation of the abatement capacity), –500, and –400 $GtCO_2$. These values are chosen to ensure clearly differentiated pathway outcomes. Under the most ambitious target (–600 $GtCO_2$), cumulative $CO_2$ emissions from 2019 onwards in the 600 $GtCO_2$ scenario are reduced to zero by 2150.

4. *T2M_Opt* Approach. This is the temperature-based approach. We use ACC2-emIAM to compute least-cost pathways that achieve a 1.5°C target by 2150, while maintaining the same emission pathways as the original IAMs up to 2100. For this approach, we imposed a constraint on temperatures to prevent further warming after 2100 and temporary emission surges. If a scenario is already below 1.5°C in 2100, the subsequent temperature trajectory is kept at that level.

Under the carbon price-based approaches (*Price_Linear* and *Price_ZeroEmi*), the abatement levels of $CH_4$ and $N_2O$ are also determined by the prescribed carbon price pathways through the respective MAC curves. Under the cumulative carbon budget-based approach (*CumBudg_Opt*), the emulator allocates $CO_2$ cost-effectively over 2100–2150, with net-zero $CO_2$ emissions by 2150. The abatement levels for $CH_4$ and $N_2O$ are subsequently estimated based on the carbon price derived from $CO_2$ mitigation pathways. In contrast, the *T2M_Opt* Approach considers trade-offs in abatement costs among $CO_2$, $CH_4$, and $N_2O$ using the GWP100 metric. In the latter two approaches, a 5% discount rate is applied for calculating the net present value of the total cost of mitigation. The different extension approaches and the corresponding results for the three gases are presented in Figs. S2 to S12.

**Funding**

This study was supported by the RESCUE project funded by the European Union's Horizon Europe research and innovation program under grant number 101056939 (W.X. and K.T.) and the OptimESM project funded by the European Union's Horizon Europe research and innovation program under grant number 101081193 (W.X. and K.T.).

## Author contributions

W.X. and K.T. conceptualized the extension framework. W.X. and K.T. simulated and analyzed the results. W.X. and K.T. co-drafted the manuscript. Both authors read and approved the final manuscript.

## Competing interests

The authors declare no competing interests.

## Supplementary Information

Supplementary information for this article is available.

## Acknowledgments

We are grateful to the two anonymous reviewers for their insightful and constructive comments, which helped improve the manuscript.

## Data availability

The datasets generated and analyzed in this study are available in the Zenodo repository, https://doi.org/10.5281/zenodo.17484824.

## Code availability

The underlying plotting code for this study is available in the Zenodo repository, https://doi.org/10.5281/zenodo.17484824. The ACC2-emIAM code is available upon the request.

*Development* **18** 1575–612

Supplement for *Extending Integrated Assessment Model scenarios until 2150 using an emulation framework*

Weiwei Xiong[1,*], Katsumasa Tanaka[1,2,*]

1 Laboratoire des Sciences du Climat et de l'Environnement (LSCE), IPSL, CEA/CNRS/UVSQ, Université Paris-Saclay, Gif-sur-Yvette, 91191, France

2 Earth System Division, National Institute for Environmental Studies (NIES), Tsukuba, 305-8506, Japan

*These authors contributed equally.

Corresponding author: Weiwei Xiong (weiwei.xiong@lsce.ipsl.fr), Katsumasa Tanaka (katsumasa.tanaka@lsce.ipsl.fr)

**Table S1 Reproducibility of emIAM for selected scenarios from each original IAM.** The table shows the consistency of three selected carbon budget scenarios between the original emissions pathways from each IAMs and calculated emissions pathways using emIAM. Cells with a deeper blue color represent higher consistency. Two indicators were used: i) ordinary Pearson's correlation coefficient $r_P$ and ii) Lin's concordance coefficient $r_C$.

| | | $rP$ | | $rC$ | |
|---|---|---|---|---|---|
| Model | Gas | With carbon price adjustment | Without carbon price adjustment | With carbon price adjustment | Without carbon price adjustment |
| AIM | $CO_2$ | 0.9563 | 0.9718 | 0.9628 | 0.9728 |
| | $CH_4$ | 0.7792 | 0.7678 | 0.7839 | 0.7852 |
| | $N_2O$ | 0.6767 | 0.6992 | 0.7197 | 0.7255 |
| COFFEE | $CO_2$ | 0.9925 | 0.9966 | 0.9926 | 0.9970 |
| | $CH_4$ | 0.9318 | 0.9601 | 0.9580 | 0.9619 |
| | $N_2O$ | 0.9328 | 0.9580 | 0.9789 | 0.9638 |
| GEM | $CO_2$ | 0.9961 | 0.9958 | 0.9964 | 0.9970 |
| | $CH_4$ | 0.9956 | 0.9988 | 0.9981 | 0.9989 |
| | $N_2O$ | 0.9070 | 0.9869 | 0.9799 | 0.9874 |
| IMAGE | $CO_2$ | 0.8324 | 0.9018 | 0.8972 | 0.9020 |
| | $CH_4$ | 0.9381 | 0.8940 | 0.9423 | 0.9129 |
| | $N_2O$ | 0.1242 | 0.4741 | 0.3619 | 0.6805 |
| MESSAGE | $CO_2$ | 0.9752 | 0.9729 | 0.9816 | 0.9786 |
| | $CH_4$ | 0.8419 | 0.8437 | 0.9545 | 0.9530 |
| | $N_2O$ | 0.9297 | 0.9551 | 0.9773 | 0.9802 |
| POLES | $CO_2$ | 0.9206 | 0.9562 | 0.9636 | 0.9584 |
| | $CH_4$ | 0.8328 | 0.8319 | 0.9710 | 0.9542 |
| | $N_2O$ | 0.8022 | 0.7657 | 0.9469 | 0.9005 |
| REMIND | $CO_2$ | 0.8640 | 0.8681 | 0.9229 | 0.9173 |
| | $CH_4$ | 0.8477 | 0.9445 | 0.9868 | 0.9564 |
| | $N_2O$ | 0.7701 | 0.8389 | 0.7987 | 0.8748 |
| TIAM | $CO_2$ | 0.9900 | 0.9943 | 0.9970 | 0.9964 |
| | $CH_4$ | 0.0114 | 0.0427 | 0.1108 | 0.2083 |
| | $N_2O$ | 0.0359 | 0.0443 | 0.6597 | 0.5540 |
| WITCH | $CO_2$ | 0.9664 | 0.9807 | 0.9856 | 0.9882 |
| | $CH_4$ | 0.7006 | 0.7510 | 0.9246 | 0.9235 |
| | $N_2O$ | 0.9740 | 0.9784 | 0.9835 | 0.9816 |

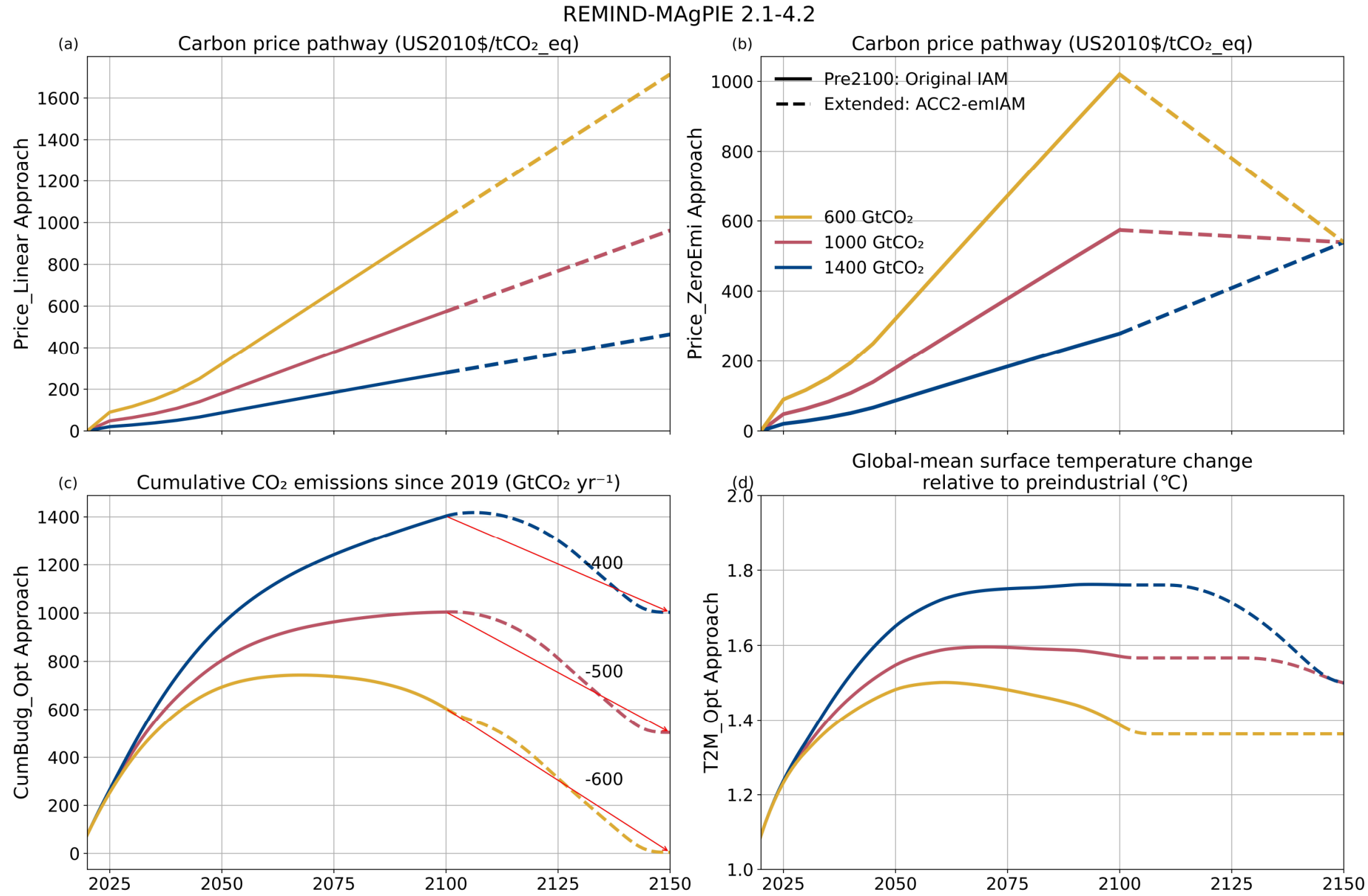

**Figure S1. Illustration of the four different Approaches to scenario extension until 2150**. Data for REMIND-MAgPIE 2.1-4.2 are used for illustration. Panels (a) and (b) present carbon price–based approaches: (a) linearly extrapolates the carbon price trend from 2090–2100 to 2150 (*Price_Linear* Approach), while (b) uses linear interpolation of carbon prices to achieve net-zero $CO_2$ emissions by 2150 (*Price_ZeroEmi* Approach). Panel (c) displays the cumulative $CO_2$ emission pathway under cumulative carbon budget constraints (*CumBudg_Opt* Approach), and (d) shows the temperature pathway consistent with the 1.5 °C target (*T2M_Opt* Approach). Both Panels (c) and (d) are generated by ACC2-emIAM with the intertemporal optimization. Solid lines represent original IAM trajectories, and dashed lines are the 2100–2150 extension.

Carbon price pathway (US2010$/tCO₂-eq)

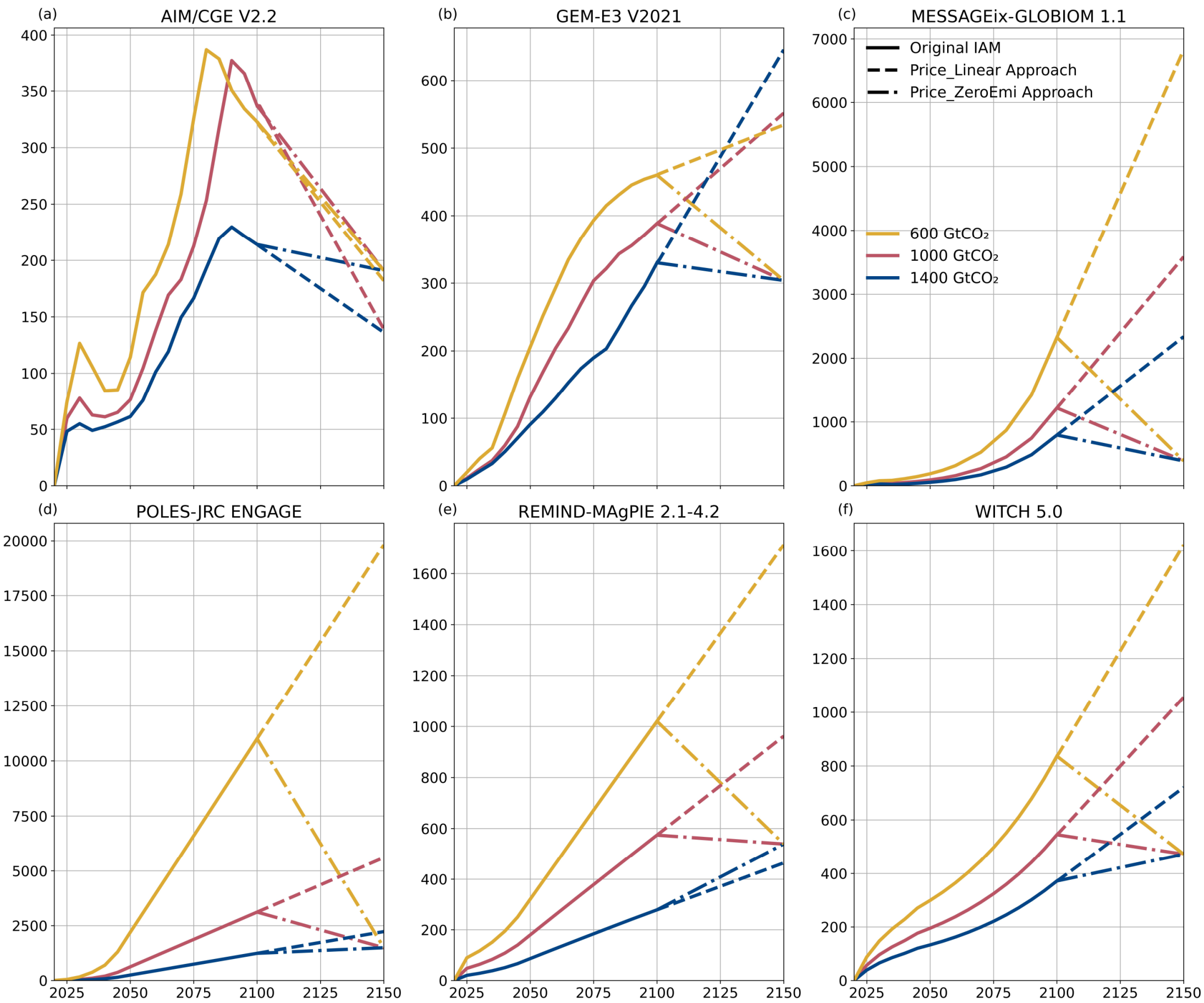

**Figure S2. Carbon price pathways developed by six IAMs with extension to 2150 using *Price-Linear* and *Price_ZeroEmi* Approaches**. Solid lines represent original IAM trajectories; dashed and dashed-dotted lines represent the constraints applied for the 2100-2150 extension using *Price-Linear* and *Price_ZeroEmi* Approaches, respectively.

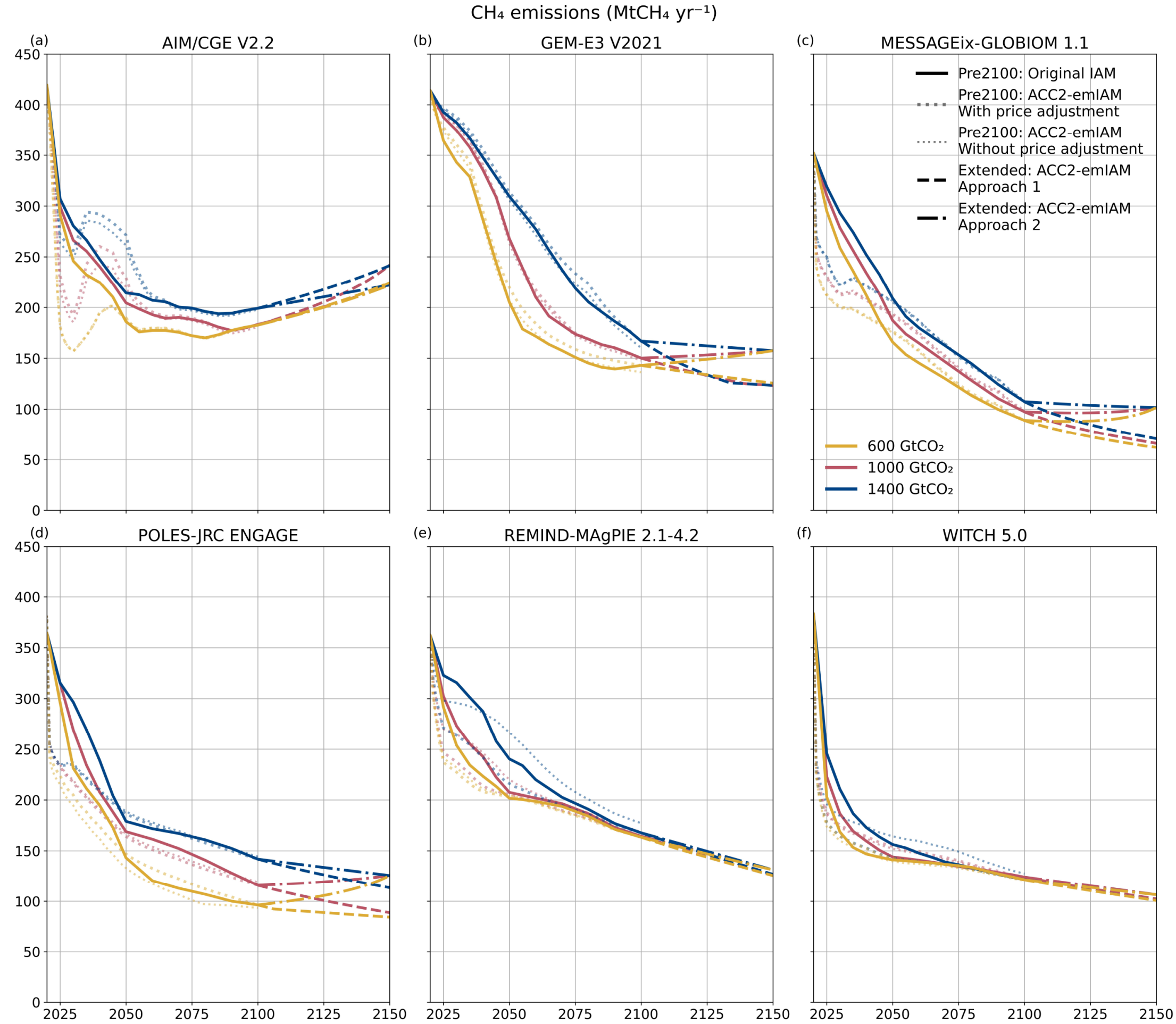

**Figure S3. $CH_4$ emission pathways developed by six IAMs with extension to 2150 using *Price_Linear* and *Price_ZeroEmi* Approaches.** Colored curves represent mitigation scenarios with different carbon budgets. Solid lines depict the original IAM results through 2100; thick and thin dotted lines show the pre-2100 results calculated by our emulator with and without the adjustment for carbon price pathways, respectively; dashed and dashed dotted lines are the 2100–2150 extension using *Price_Linear* and *Price_ZeroEmi* Approaches, respectively.

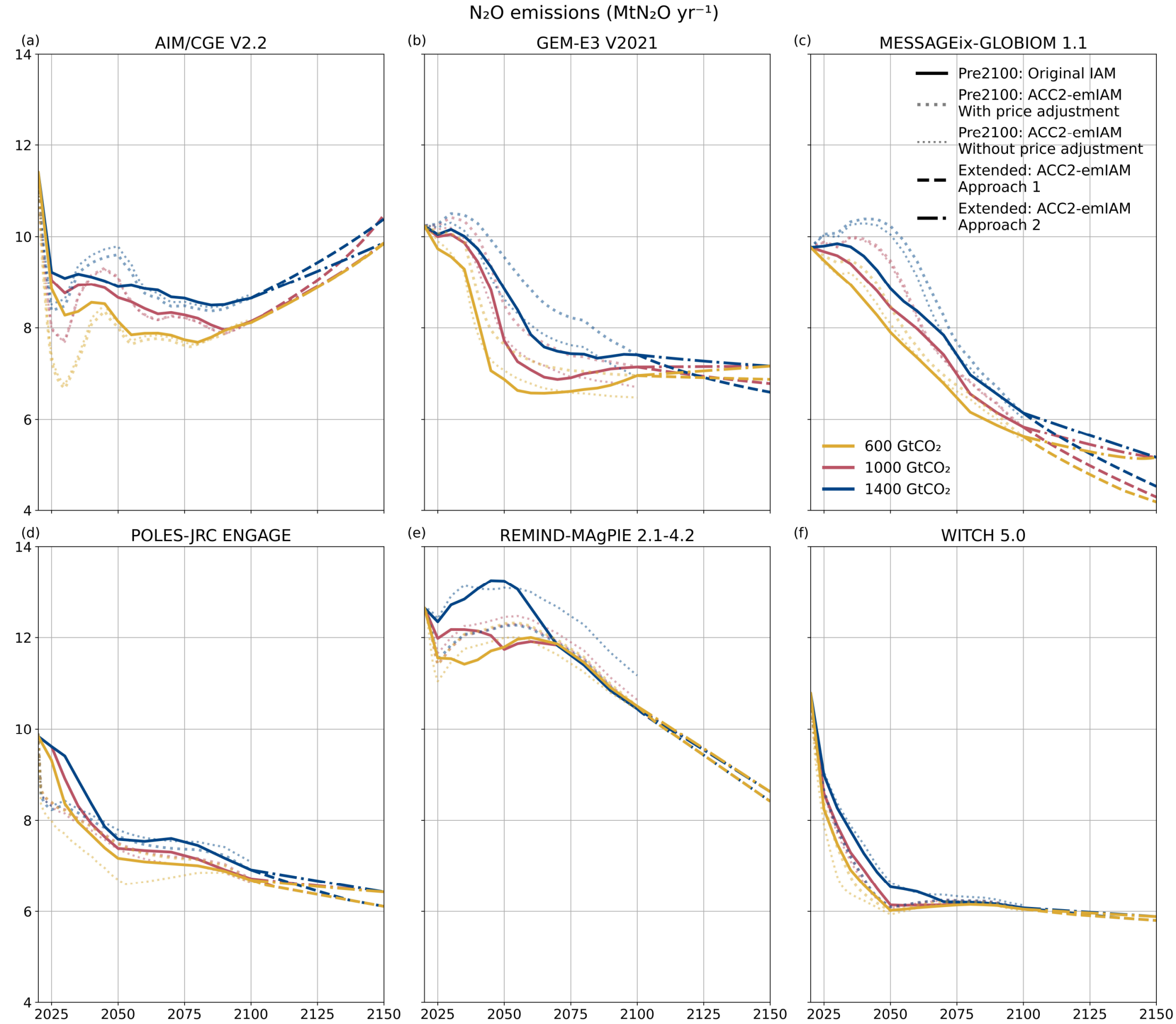

**Figure S4. $N_2O$ emission pathways developed by six IAMs with extension to 2150 using *Price_Linear* and *Price_ZeroEmi* Approaches.** Colored curves represent mitigation scenarios with different carbon budgets. Solid lines depict the original IAM results through 2100; thick and thin dotted lines show the pre-2100 results calculated by our emulator with and without the adjustment for carbon price pathways, respectively; dashed and dashed dotted lines are the 2100–2150 extension using *Price_Linear* and *Price_ZeroEmi* Approaches, respectively.

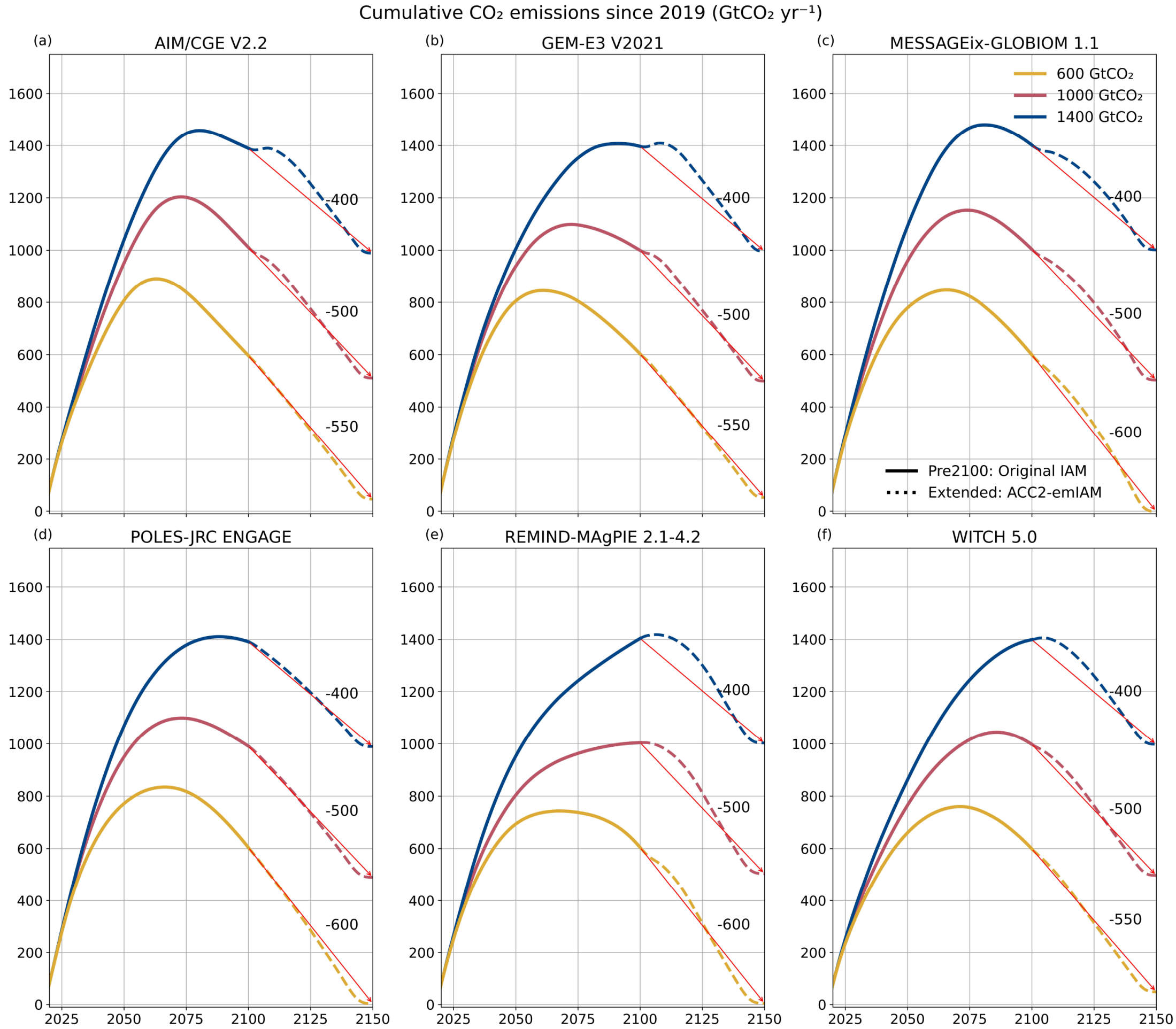

**Figure S5. Cumulative $CO_2$ emission pathway developed by six IAMs with extension to 2150 using the *Cumbudg_Opt* Approach**. Solid lines represent original IAM trajectories, and dashed lines denote the constraints applied for the 2100-2150 extension.

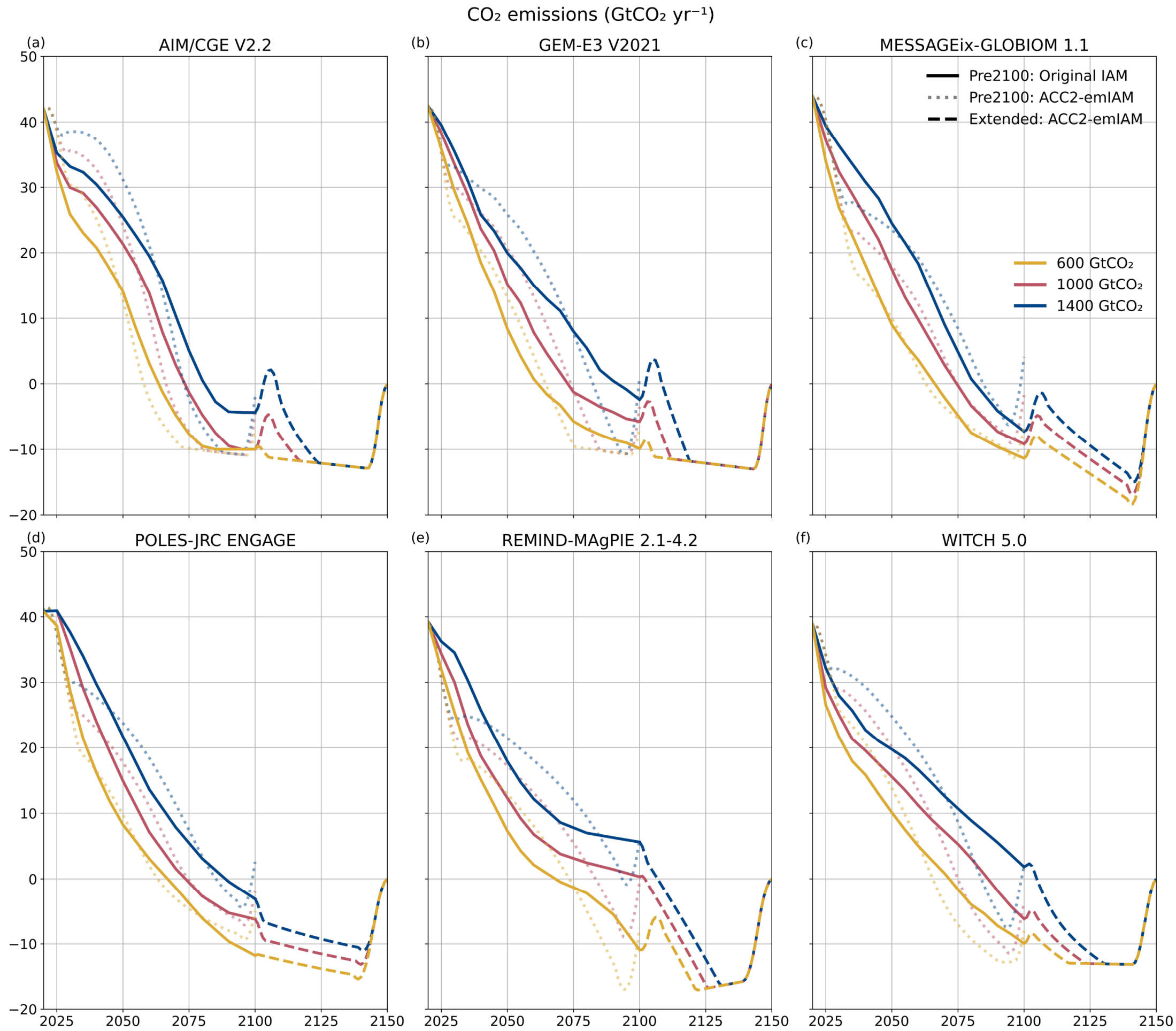

**Figure S6. $CO_2$ emission pathways developed by six IAMs with extension to 2150 using the *Cumbudg_Opt* Approach.** Colored curves represent mitigation scenarios with different carbon budgets. Solid lines depict the original IAM results through 2100; dotted lines show the pre-2100 results calculated by our emulator; dashed lines are the 2100–2150 extension.

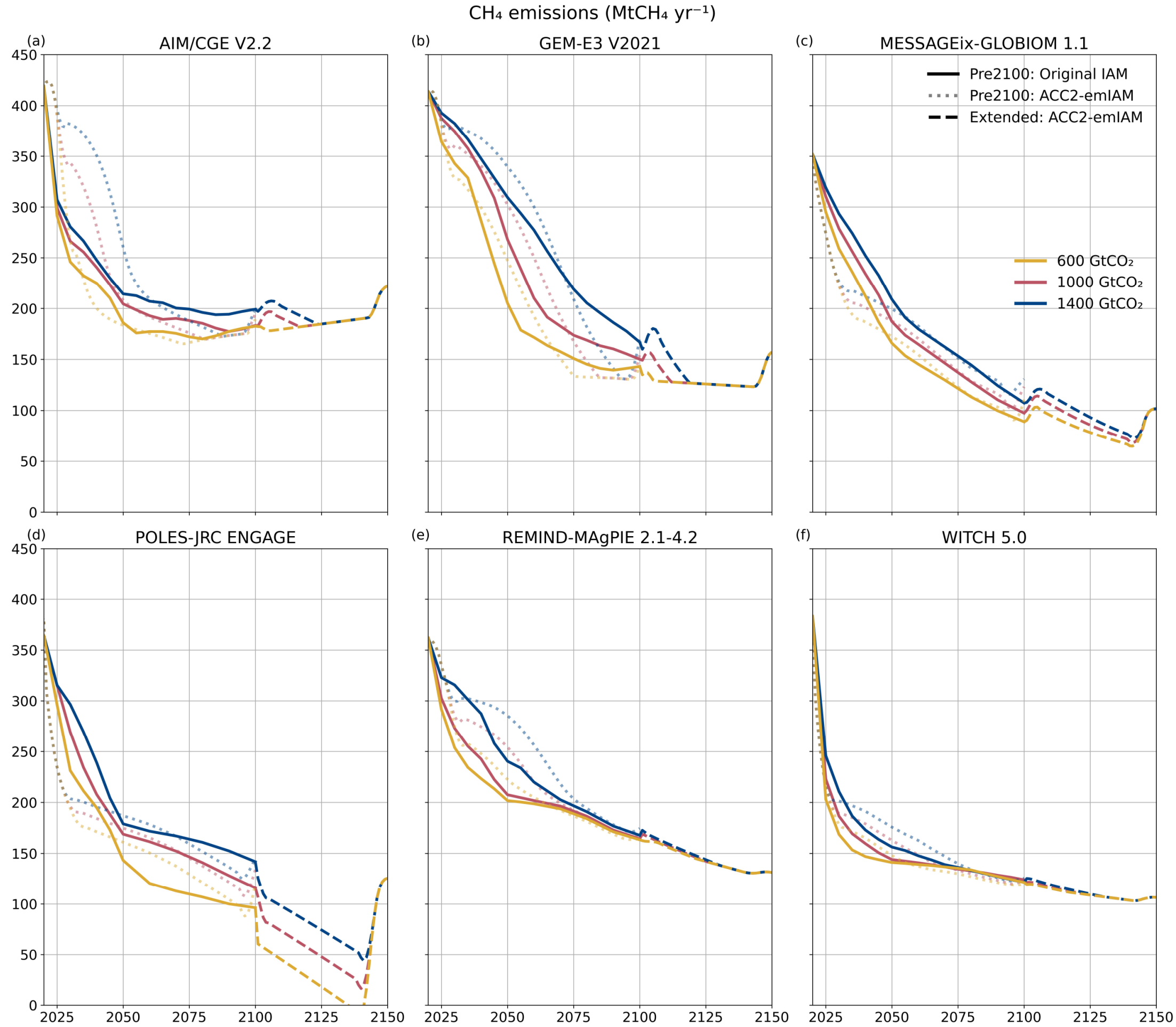

**Figure S7. $CH_4$ emission pathways developed by six IAMs with extension to 2150 using the *Cumbudg_Opt* Approach.** Colored curves represent mitigation scenarios with different carbon budgets. Solid lines depict the original IAM results through 2100; dotted lines show the pre-2100 results calculated by our emulator; dashed lines are the 2100–2150 extension.

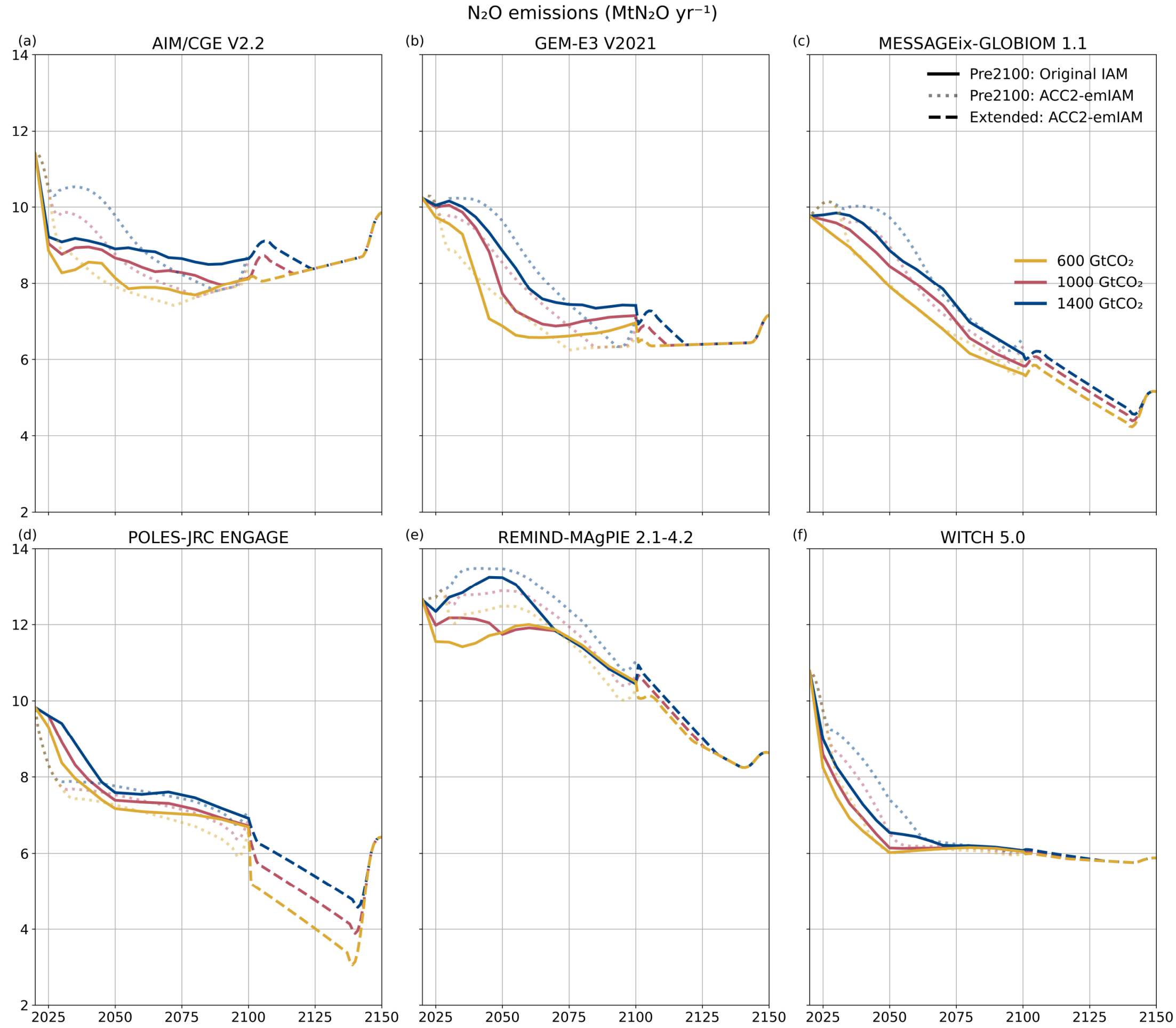

**Figure S8. $N_2O$ emission pathways developed by six IAMs with extension to 2150 using the *Cumbudg_Opt* Approach.** Colored curves represent mitigation scenarios with different carbon budgets. Solid lines depict the original IAM results through 2100; dotted lines show the pre-2100 results calculated by our emulator; dashed lines are the 2100–2150 extension.

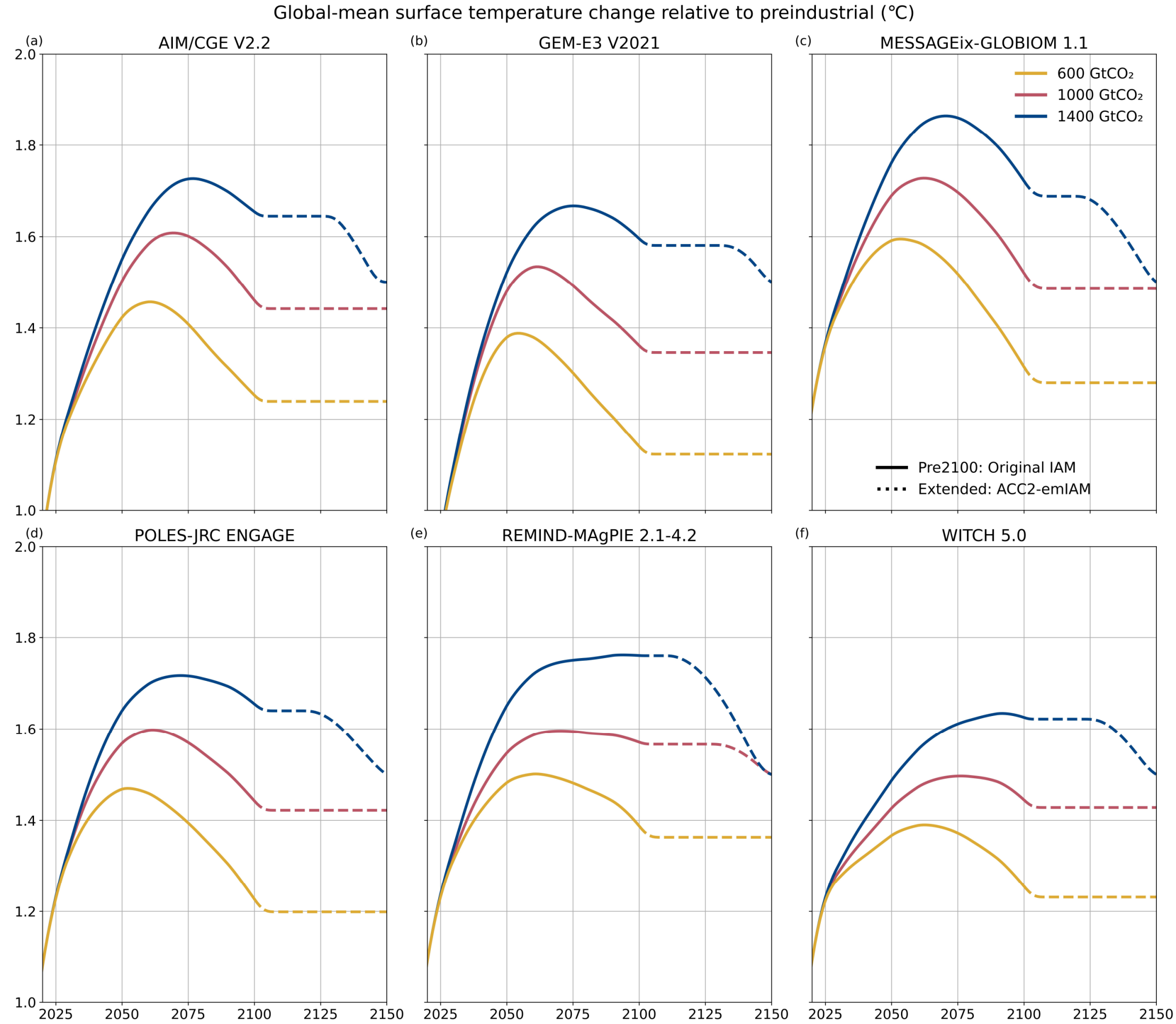

**Figure S9. Temperature pathways developed by six IAMs with extension to 2150 using the *T2M_Opt* Approach**. Solid lines represent original IAM trajectories calculated by ACC2; dashed lines denote the constraints applied for the 2100-2150 extension, which are generated by ACC2-emIAM with the intertemporal least-cost optimization.

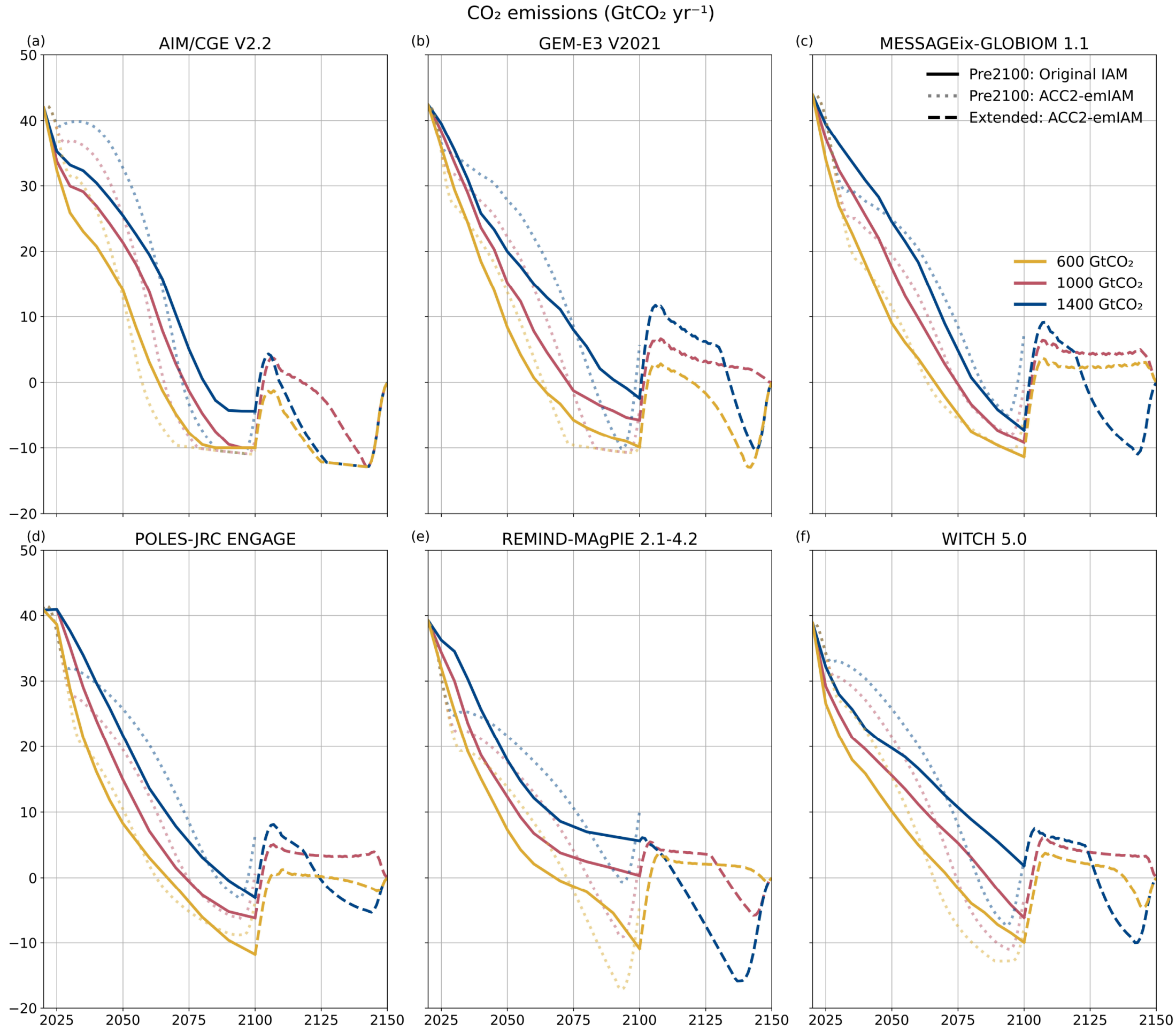

**Figure S10. $CO_2$ emission pathways developed by six IAMs with extension to 2150 using the *T2M_Opt* Approach.** Colored curves represent mitigation scenarios with different carbon budgets. Solid lines depict the original IAM results through 2100; dotted lines show the pre-2100 results calculated by our emulator; dashed lines are the 2100–2150 extension.

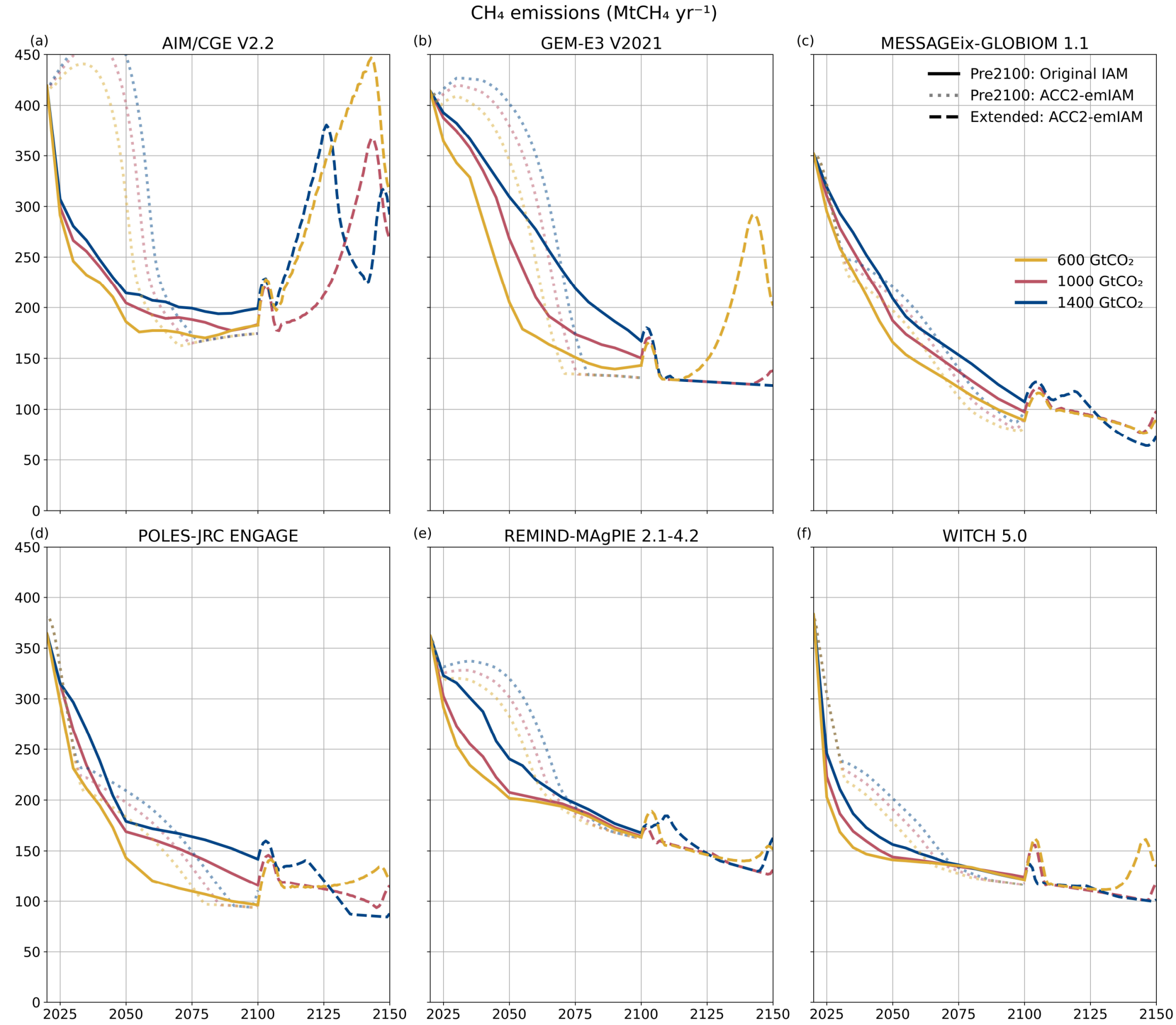

**Figure S11. $CH_4$ emission pathways developed by six IAMs with extension to 2150 using the *T2M_Opt* Approach.** Colored curves represent mitigation scenarios with different carbon budgets. Solid lines depict the original IAM results through 2100; dotted lines show the pre-2100 results calculated by our emulator; dashed lines are the 2100–2150 extension.

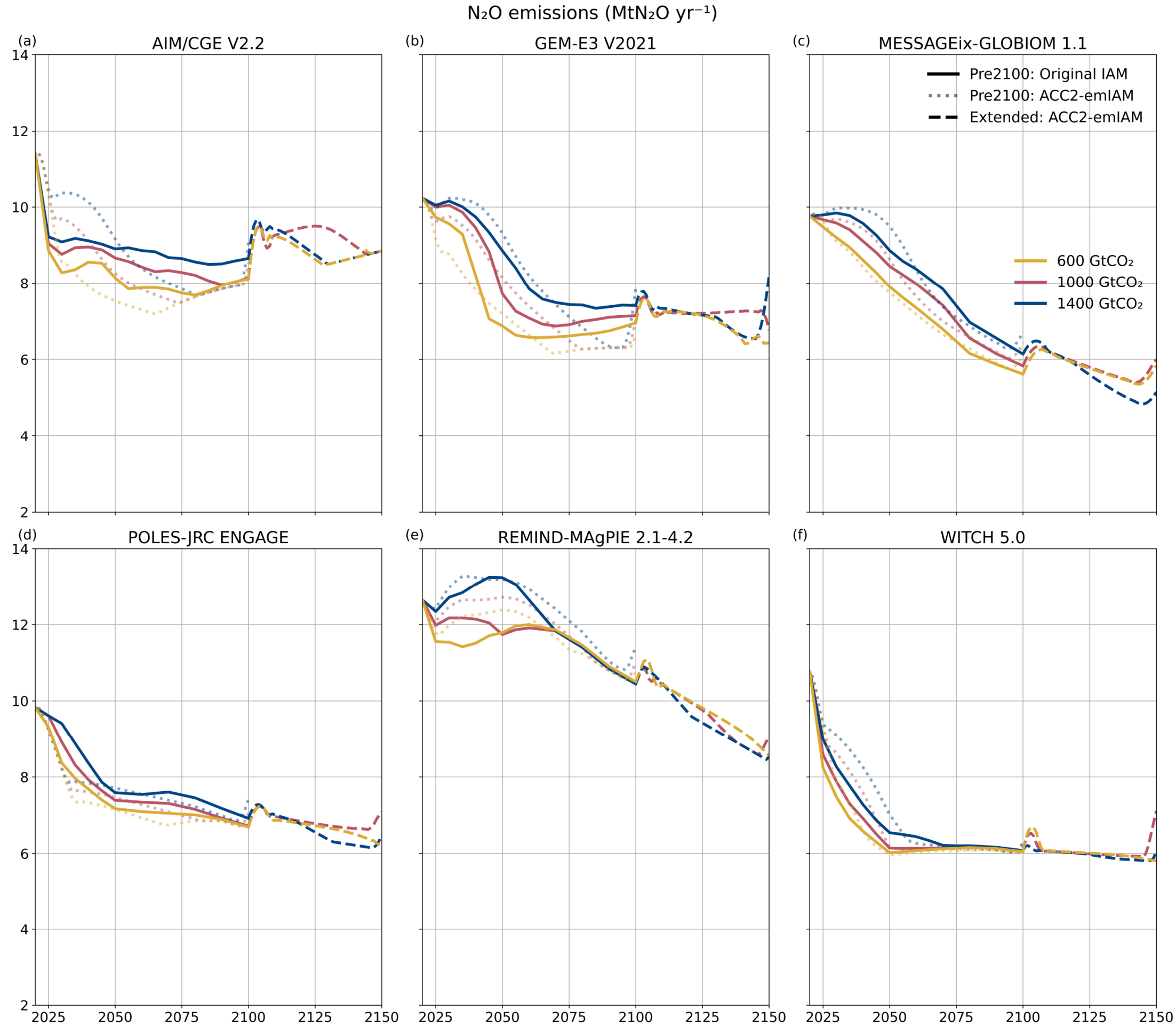

**Figure S12. $N_2O$ emission pathways developed by six IAMs with extension to 2150 using the *T2M_Opt* Approach.** Colored curves represent mitigation scenarios with different carbon budgets. Solid lines depict the original IAM results through 2100; dotted lines show the pre-2100 results calculated by our emulator; dashed lines are the 2100–2150 extension.